\documentclass{article}
\usepackage{spconf,amsmath,graphicx}
\usepackage{url}
\usepackage{enumitem}
\setlist{nosep, leftmargin=14pt}
\usepackage{multirow}
\usepackage{mwe} 

\title{QUANTITATIVE METRICS FOR BENCHMARKING MEDICAL IMAGE HARMONIZATION}
\name{\parbox{\linewidth}{\centering
Abhijeet Parida\textsuperscript{1},
Zhifan Jiang\textsuperscript{1},
Roger J. Packer\textsuperscript{1},
Robert A. Avery\textsuperscript{2},
Syed M. Anwar\textsuperscript{1,3},
Marius G. Linguraru\textsuperscript{1,3}
}}

\address{\parbox{\linewidth}{\centering\fontsize{9}{10}\selectfont
\textsuperscript{1} Children’s National Hospital, Washington, DC, USA \\
\textsuperscript{2} Children’s Hospital of Philadelphia, Philadelphia, PA, USA \\
\textsuperscript{3} School of Medicine and Health Sciences, George Washington University, Washington, DC, USA
}}
\begin{document}
%
\maketitle
\begin{abstract}
Image harmonization is an important preprocessing strategy to address domain shifts arising from data acquired using different machines and scanning protocols in medical imaging. However, benchmarking the effectiveness of harmonization techniques has been a challenge due to the lack of widely available standardized datasets with ground truths. In this context, we propose three metrics- two intensity harmonization metrics and one anatomy preservation metric for medical images during harmonization, where no ground truths are required.  Through extensive studies on a dataset with available harmonization ground truth, we demonstrate that our metrics are correlated with established image quality assessment metrics.  We show how these novel metrics may be applied to real-world scenarios where no harmonization ground truth exists. Additionally, we provide insights into different interpretations of the metric values, shedding light on their significance in the context of the harmonization process. As a result of our findings, we advocate for the adoption of these quantitative harmonization metrics as a standard for benchmarking the performance of image harmonization techniques.
\end{abstract}

\section{Introduction}
\label{sec:intro}
\vspace{-0.2cm}
Deep learning (DL) models, trained on extensive, well-curated radiologic data from similar sources, are effective in aiding disease diagnosis and clinical trials.  When making clinical comparisons or contributing data to clinical trials, a variety of imaging equipment brands and acquisition protocols are utilized both within and across radiology departments. This leads to a wide range of variations in imaging data and visual appearance for a similar clinical condition, especially in magnetic resonance imaging (MRI), which lacks a standardized unit. These data distribution variations or lack of harmonization across studies can lead to an increase in error in clinical tasks performed using DL algorithms trained with non-harmonized data. 

To address the challenges of domain shifts inherent in MRI, most computational methods employ image harmonization as a preprocessing step before performing an image analysis task (such as classification or segmentation). Most DL methods for medical image harmonization are based on generative models (such as generative adversarial networks [GANs]) that utilize adversarial loss functions \cite{mirza2014conditional, sinha2021alzheimer, medstyle, liu_unit_2017, carlos}. However, these GAN-based methods are known to generate structures that may not be present in the original image, a phenomenon known as hallucination, which negatively impacts the clinical interpretation of the image \cite{yang_cyclegan_2018}. It is important to measure these harmonization artifacts, but dedicated metrics are lacking. Hence, the field of image harmonization would benefit from quantitative metrics that can be used to benchmark the quality of harmonization. 

Image generation metrics have been used to determine image harmonization quality. These metrics fall into two categories: image-level and feature-level metrics \cite{harm_review}. Image-level metrics, such as mean absolute error (MAE), mean squared error (MSE), or peak signal-to-noise ratio (PSNR) \cite{psnr}, offer high-quality measurements but require paired harmonized data, which can be challenging to obtain \cite{denck2021mr}. Metrics like structural similarity (SSIM) \cite{ssim} can be used to alleviate the constraint on paired data \cite{sinha2021alzheimer}, but they primarily compare structures at a higher level and could miss smaller hallucinations generated by GANs, which are clinically relevant. 
Feature-level metrics, such as t-distributed stochastic neighbor embedding (t-SNE) for visualizing the manifold, can be utilized to evaluate harmonization performance, as demonstrated in \cite{medstyle}. Other techniques, like principal components analysis and uniform manifold approximation and projection \cite{umap}, can be similarly applied to visually assess the effectiveness of harmonization. However, it is worth noting that these metrics require dimensionality reduction, which results in information loss and neglect of spatial relationships that may be critical for medical applications. 

Another approach to assess harmonization is to employ task-specific evaluations, such as measuring cortical thickness for brains  \cite{medstyle}, segmenting specific anatomical structures \cite{carlos}, or conducting disease classification \cite{sinha2021alzheimer}. These methods are tailored to a clinically relevant task, where the goal of image harmonization is to prepare images to enhance their consistency and suitability for that specific task only. Task-specific evaluations offer a limited perspective of the quality of the harmonization, and the metrics used may not be directly comparable across different tasks. For instance, harmonization may yield excellent performance for one type of brain tumor but exhibit lower performance for other types of tumors due to the induced task-specific bias. 

To address the limitations of the current standard metrics, we present new metrics that can be used to quantify the performance of medical image harmonization methods. Our major contributions include: 1) introducing novel interpretable metrics to measure the intensity harmonization and anatomical changes in medical images; 2) conducting a detailed analysis of brain MRI harmonization utilizing these new metrics on data collected from sites using variable imaging devices and protocols, and 3) comparing the new metrics against established metrics for a comprehensive evaluation.

\section{Evaluation of Image Harmonization }\label{me}
\vspace{-0.2cm}
We argue that a robust evaluation strategy for medical imaging harmonization should assess: 1) intensity harmonization, i.e., the appearance of the predicted/harmonized image matches that of the target image/protocol; 2) anatomy preservation, i.e., the anatomical structures should not change through the process of harmonization. This process would promote evaluation interpretation agnostic to the clinical or machine learning task for which harmonization is performed. Therefore, the harmonization metrics would help the consumer make informed decisions when training DL methods. 
\vspace{-0.5cm}
\subsection{Intensity Harmonization}

We propose using the Wasserstein distance (WD) \cite{wasser} to evaluate the intensity harmonization by measuring the movement of intensity histograms. We chose WD over Jensen-Shannon (JS) or Kullback-Leibler (KL) divergences because Jensen-Shannon is a fixed value for non-overlapping distributions, and Kullback-Leibler is not defined for non-overlapping distributions \cite{kolouri2018sliced}. These limitations of KL and JS could lead to unreliable metrics when the intensity distributions of images at different sites are highly dissimilar and non-overlapping. 

\begin{figure}[h]
    \centering
    \includegraphics[width=\columnwidth]{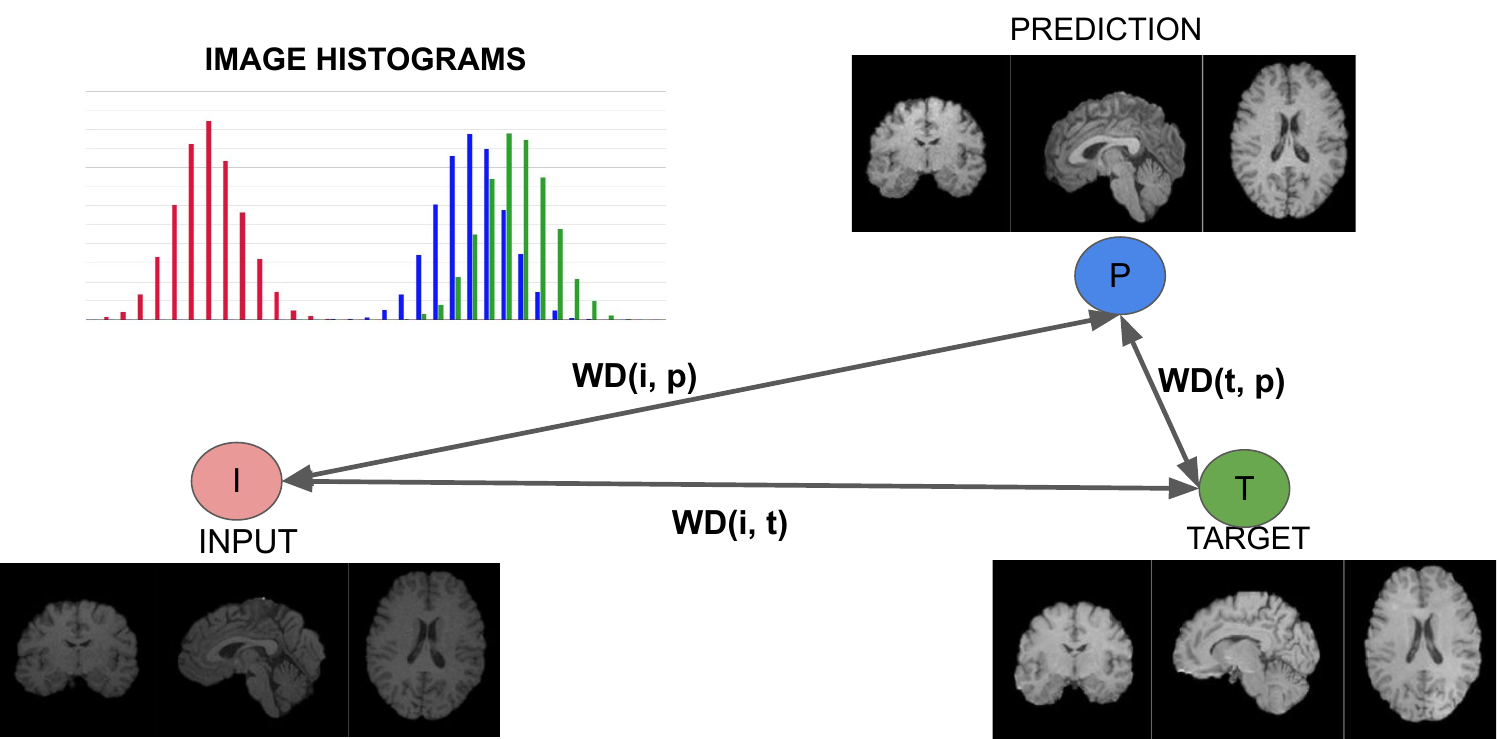}
    \abovecaptionskip=3pt
    \belowcaptionskip=3pt
    \caption{\textbf{Image Harmonization:} Schematic of the image intensity histogram manifold and how the various metrics (WD(i,p), WD(t, p), and WD(i, t)), with the input image(i), target(t), and the predicted image(p). }\vspace{-0.3cm}
    \label{fig:wasser}
\end{figure}

We define $WD(i,t)$ as the distance between input ($i$) and target ($t$) images as the upper bound for the model prediction performance, as shown in Figure \ref{fig:wasser}. We also define $WD(t,p)$ as the distance between $t$ and prediction ($p$). To make the metric scale agnostic and comparable across data from different sites, we define $nWD$, which is the normalized metric by $WD(i,t)$, i.e.,  
\begin{equation}
\begin{split}
 nWD(i,p)=\frac{WD(i,p)}{WD(i,t)}, \\ 
 nWD(t,p)=\frac{WD(t,p)}{WD(i,t)}. 
 \end{split}
\end{equation}
For superior performance in intensity harmonization, we anticipate a high $nWD(i,p)$ value and a small $nWD(t,p)$ value. Some additional interpretations of the metrics include:
\begin{itemize}
  \item when $nWD(i, p) = 0\ \&\ nWD(t, p) = 1$ means no intensity harmonization was performed,
  \item when $nWD(i, p) = 1\  \&\ nWD(t, p) = 0$ represents perfect intensity harmonization,
  \item when $nWD(i, p) > 1$ is a sign of over-correction of the image intensities. 
\end{itemize}
\vspace{-0.4cm}
\vspace{-0.3cm}
\subsection{Anatomy Preservation}
\vspace{-0.6cm}
\begin{figure}[h!]
    \centering
    \includegraphics[width=0.90\columnwidth]{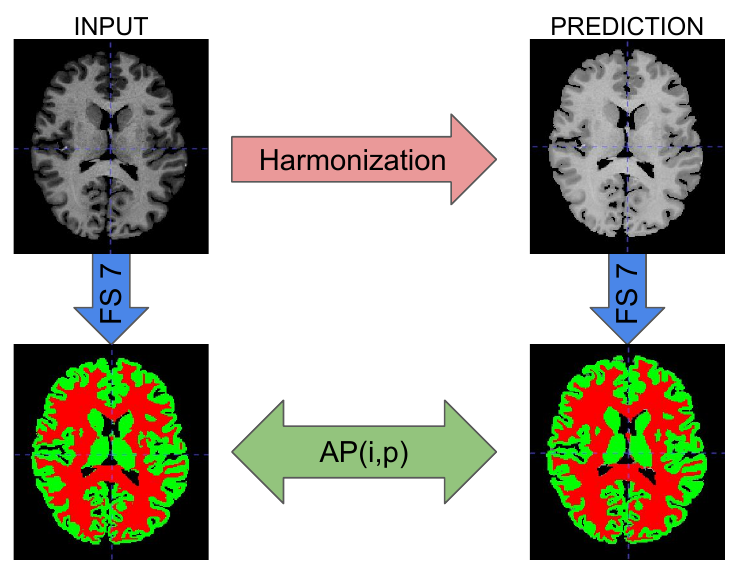}
    \abovecaptionskip=3pt
    \belowcaptionskip=3pt
    \caption{\textbf{Anatomy Preservation:} Schematic of AP(i,p) calculation for brain image harmonization using Freesurfer v7 (FS7).}\vspace{-0.3cm}
    \label{fig:fs7}
\end{figure}

We propose using a method that automatically segments anatomical structures in the input and the predicted image and then uses an established metric for comparison. This approach guarantees that harmonized outputs are suitable for downstream tasks, such as segmentation. We propose using anatomy preservation $(AP)$ to measure the conservation of the relative absolute volume of anatomical structures segmented before and after data harmonization. We define AP to be agnostic to the size of the anatomical structures as

\begin{equation}
 AP(i,p)= 1 - \frac{|vol(p)-vol(i)|}{vol(i)},  
\end{equation}
where $vol(i)$ and $vol(p)$ denote input and predicted harmonized volumes of the anatomical structure. For good performance in anatomy preservation, we expect a large $AP(i,p)$ value, ideally 1 meaning all the anatomy is preserved. 

We use the established Freesurfer v7 software \cite{fischl2012freesurfer} to segment the brain gray matter (GM) and white matter (WM) regions to calculate $AP(i,p)$, of the example of brain MRI  harmonization that we show in the paper. Freesurfer models have been trained on large datasets and are robust to a wide range of data shifts\cite{zavaliangos2022testing}. Also importantly, they are publicly available to the scientific community. Finally, we report $AP(i,p)$, which is the average of the $AP(i,p)_{GM}$ and $AP(i,p)_{WM}$.

\section{Experimental Setup}
\label{sec:res}
\vspace{-0.2cm}
To assess the effectiveness of our metrics, we conducted experiments using the GAN-based neural style transfer (NST) brain image harmonization method proposed in \cite{medstyle} in two distinct scenarios: 1) Ground truth evaluation based on a traveling human phantom dataset with paired data, and 2) Real-world scenario with brain MRI from different sites where paired data is unavailable. 
\vspace{-0.3cm}
\subsection{Ground Truth Evaluation}
\noindent
\textbf{Dataset:} 
This evaluation emulates ideal conditions where ground truth from paired data is available. We utilized the frequently traveling human phantom dataset \cite{opfer2023automatic}, which comprises 557 3D T1-weighted gradient echo sequences from a 50-year-old male subject. Notably, this individual underwent 157 imaging sessions across 116 different MRI scanners spanning five imaging sites. We processed the data to reflect the site-specific structure based on the ’SiteID’ and ’ScannerID’. The processed data has been made accessible on Kaggle\footnote{\url{https://www.kaggle.com/datasets/aparida/sitewise-frequently-traveling-human-phantom/}} for the benefit of the wider research community. 

\noindent
\textbf{Data Processing:} 
All brain MRIs were skull-stripped and registered to the MNI152 atlas \cite{mnitemplate} following the procedures outlined in \cite{medstyle}. 

\noindent
\textbf{Experiment Design:} 
We selected only sites with over 10 images, resulting in 3 sites. For each of these target sites, we used the GAN-based NST framework \cite{medstyle} and a randomly chosen image (i.e., $049.nii$) was harmonized for each target site using the pre-trained NST weights provided by \cite{medstyle}. 

\noindent
\textbf{Metrics:} 
To facilitate a comprehensive comparison between our proposed metrics and established image-level metrics such as MAE, MSE, SSIM, and PSNR, we employed pairs of harmonized images, i.e., the predicted image and its corresponding ground truth image. All the metrics are calculated after image background removal to avoid the averaging effects of a large background in the image.
\vspace{-0.3cm}
\subsection{Real-world Scenario}

\noindent
\textbf{Dataset:} 
This scenario emulates real-world conditions where ground truth data is unavailable. We collected pediatric brain MRI data from two sites: Children’s National Hospital (Site A) and Children’s Hospital of Philadelphia (Site B). Each site contributed n = 60 3D T1-weighted MRIs, each acquired using varying acquisition protocols and scanners, as outlined in Table \ref{tab:data}. 

\begin{table}[h!]
{\small
\centering
   \begin{tabular}{|c|c|c|}
\hline
& \textbf{Site A} & \textbf{Site B} \\ \hline\hline
\textbf{\begin{tabular}[c]{@{}c@{}}MRI Manufacturer\end{tabular}}                           & \begin{tabular}[c]{@{}c@{}}GE\end{tabular} & Siemens         \\ \hline
\textbf{\begin{tabular}[c]{@{}c@{}}Acquisition Plane\end{tabular}}                           & Axial                                                      & Sagittal \\ \hline
\textbf{TE/TR (ms)}                                                                             & 10.5/600                                                   & 2.5/1900       \\ \hline
\textbf{\begin{tabular}[c]{@{}c@{}}Resolution ($\boldsymbol{mm^3})$\end{tabular}} & $0.41\times0.41\times0.6$                                                  & $0.82\times0.82\times0.9$       \\ \hline
\end{tabular}
\caption{\textbf{Dataset summary} of the different acquisition protocols of the pediatric brain MRIs at each of the sites A and B. TR: repetition time, TE: echo time.}\vspace{-0.3cm}
\label{tab:data}
}
\end{table}

\noindent
\textbf{Data Processing:} 
Due to computational resource limitations and our exclusive focus on intensity harmonization, we adjusted the MRI resolution to $1\times1\times1 ;mm^3$ and resized images to $256\times256\times256$ voxels.

\noindent
\textbf{Experiment Design:} 
We retrained the GAN-based NST method for 1,000,000 iterations, with a batch size 8, and learning rate of $1e^{-4}$ using the Adam optimizer as detailed in \cite{medstyle}.

\noindent
\textbf{Metrics:} 
Given the absence of ground truth data, we exclusively report the proposed intensity harmonization metrics, i.e., $nWD(i,p)$ and $nWD(i,t)$, and the anatomy preservation metric, i.e., $AP(i,p)$. 

\section{Results \& Interpretations} \label{sec:dis}
\vspace{-0.3cm}
\subsection{Ground Truth Evaluation and Comparison with Other Metrics }
We conducted a comprehensive comparison of all the metrics across all scans and present the results in Table \ref{tab:res2}. Based on the SSIM, PSNR, MAE, and MSE values, it is evident that the harmonization is not very good with the most effective harmonization among the three being achieved for site 1, while the least effective harmonization is observed for site 3. The similar effect can also be interpreted with the lower values of $nWD(i,p)$ and the high values of $nWD(i,t)$ as the intensity ranges of the brain scans are insufficiently far from the input site and not very close to the target site. 
\begin{table*}[]
\centering
{
\resizebox{.99\textwidth}{!}{%
\begin{tabular}{|c|c|c|c|c|c|c|c|}
\hline
  & \textbf{SSIM}       & \textbf{PSNR}    & \textbf{MAE}        & \textbf{MSE}        & \textbf{nWD(i, p)}  & \textbf{nWD(t, p)}  & \textbf{AP(i,p)} \\ \hline\hline
\textbf{Site 1} & 0.599$\pm$ 0.050  & 17.351 $\pm$ 1.479 & 0.0719 $\pm$ 0.013 & 0.019 $\pm$ 0.006 & 0.468 $\pm$ 0.122 & 1.366 $\pm$ 0.053  & 0.974   $\pm$ 0.012         \\ \hline
\textbf{Site 2} & 0.422 $\pm$ 0.092 & 14.057 $\pm$ 0.977 & 0.113 $\pm$ 0.020 & 0.040 $\pm$ 0.009 & 0.199 $\pm$ 0.100    & 0.959 $\pm$ 0.057 & 0.961  $\pm$   0.024        \\ \hline
\textbf{Site 3} & 0.347 $\pm$ 0.096 & 13.744 $\pm$ 1.896 & 0.131 $\pm$ 0.036 & 0.045 $\pm$ 0.019 & 0.197 $\pm$ 0.150 & 0.999 $\pm$ 0.108 & 0.960  $\pm$    0.015         \\ \hline
\end{tabular}%
}
}
\caption{\textbf{Comparative analysis} of our metrics with standard metrics for image harmonization on the traveling human phantom data set.}\vspace{-0.3cm}
\label{tab:res2}
\end{table*}
Furthermore, by examining the $nWD(i,p)$  values, it is apparent that the performance of intensity harmonization is superior for site 1 compared to site 2 and site 3. For sites 2 and 3, interpreting the $nWD(t,p)$ values, $nWD(t,p) \rightarrow 1$ with low $nWD(i,p)$ means actually not much harmonization was achieved. 

The $AP(i,p)$ values indicate that there are small changes in the volume of the GM and WM before and after the harmonization process for the sample $049.nii$. 

In Table \ref{tab:corr}, we present Spearman rank correlation values among the metrics. Metrics evaluating positive image attributes, such as SSIM and PSNR for assessing image generation quality, show positive correlations with the proposed metrics $nWD(i,p)$ and $AP(i,p)$. Conversely, metrics related to negative image attributes, like MSE and MAE, serving as error measures, demonstrate negative correlations with the proposed metrics. This implies that $nWD(i,p)$ and $AP(i,p)$ can serve as substitutes for metrics like SSIM, PSNR, MSE, and MAE, especially in scenarios where their calculation is not feasible due to the absence of ground truth, as is often the case in many harmonization tasks. 
\begin{table}[h!]
\centering
{
\begin{tabular}{|l|l|l|l|l|}
\hline
          & \textbf{SSIM} & \textbf{PSNR} & \textbf{MAE} & \textbf{MSE} \\ \hline\hline
\textbf{nWD(i,p)}  &  0.988    &    0.997  & -0.965    & -0.990    \\ \hline
\textbf{nWD(t,p)}  &  -0.482    &    -0.525 & 0.071    & 0.011   \\ \hline
\textbf{AP(i,p)} &   0.967   &  0.999    &  -0.964   &  -0.986   \\ \hline
\end{tabular}
}

\caption{\textbf{Spearman rank correlation} between metrics for the traveling human phantom dataset.}\vspace{-0.3cm}
\label{tab:corr}
\end{table}

Furthermore in Table \ref{tab:corr}, $nWD(t,p)$ shows a negative correlation with SSIM and PSNR, and small positive correlations with MSE and MAE. Thus, it serves as a good indicator of errors, with higher $nWD(t,p)$ values indicating lower harmonization performance.

This observation suggests that reporting the metrics $nWD(i,p)$, $nWD(t, p)$, and $AP(i,p)$ together can provide us with an estimate of the quality of harmonization as well as the degree of deviation in the generated harmonized brain compared to the input brain image.
\vspace{-0.3cm}
\subsection{Performance on Real World Data}
\vspace{-0.5cm}
\begin{figure}[h]
    \centering
    \includegraphics[width=0.95\columnwidth]{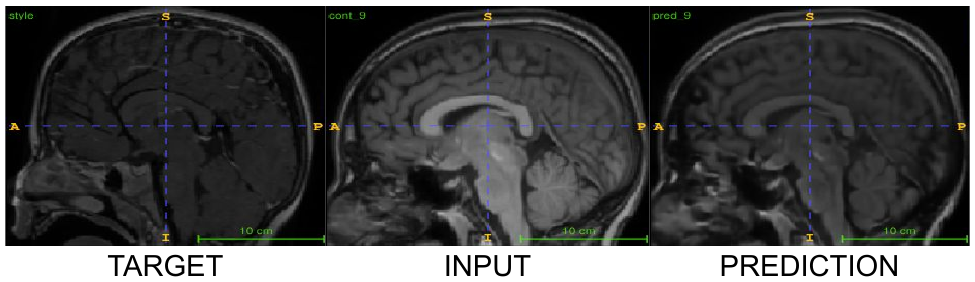}
    \abovecaptionskip=3pt
    \caption{\textbf{Qualitative results} show harmonization of images from Site A(input) to B(target) and the predicted harmonized image.}\vspace{-0.5cm}
    \label{fig:harm}
\end{figure}
In Table \ref{tab:res1}, we present the model’s performance in harmonizing data from site A to site B and vice versa. The values of $nWD(i, p)$ and $nWD(t, p)$ collectively indicate that the intensity values of the predicted image are further from the input images and much closer to the target images. This can also be visually confirmed in Figure 3. In contrast to the results in Table \ref{tab:res2}, Table \ref{tab:res1} reveals that retraining the harmonization model specifically for the dataset resulted in improved harmonization, i.e., high $nWD(i,p)$ and low $nWD(t, p)$, without over-correction, as indicated by $nWD(i, p) < 1$. 

\begin{table}[h!]
\centering
{
\resizebox{.99\columnwidth}{!}{
\begin{tabular}{|c|c|c|c|}
\hline
\multirow{2}{*}{\textbf{}} & \multicolumn{2}{c|}{\textbf{\begin{tabular}[c]{@{}c@{}}Intensity\\ Harmonization\end{tabular}}} & \multicolumn{1}{c|}{\textbf{\begin{tabular}[c]{@{}c@{}}Anatomy \\ Preservation\end{tabular}}} \\ \cline{2-4} 
                           & \multicolumn{1}{c|}{\textbf{nWD(i, p)}}                 & \textbf{nWD(t, p)}                & \textbf{AP(i, p)}                                                                         \\ \hline\hline
\textbf{A $\rightarrow$ B}               & \multicolumn{1}{l|}{0.906 $\pm$ 0.038}                           & 0.087 $\pm$ 0.006                           & 0.899 $\pm$ 0.193                                                                                   \\ \hline
\textbf{B  $\rightarrow$ A}               & \multicolumn{1}{l|}{0.967 $\pm$ 0.033}                           & 0.074 $\pm$ 0.006                           & 0.937 $\pm$ 0.047                                                                                    \\ \hline
\end{tabular}
}
}

\caption{\textbf{Quantitative Results} for image harmonization between two sites. The higher $nWD(i,p)$ compared to the $nWD(t,p)$ indicates a good transfer of intensities. The high $AP(i,p)$ suggests that the anatomies are preserved during harmonization. }\vspace{-0.3cm}
\label{tab:res1}
\end{table}

Figure \ref{fig:harm}  illustrates that the structures between the input and predicted images are well preserved, and the preservation of anatomical structures is quantified by $AP(i, p)$ in Table \ref{tab:res1}.

\vspace{-0.2cm}
\section{Discussions}
\vspace{-0.2cm}
\label{sec:conclu}
 These metrics are independent of each other and have an interpretable meaning in terms of harmonization. However, we demonstrate the effective utilization of the $nWD(i,p)$ and $nWD(t,p)$ metrics for 3D single-channel images. These metrics can be seamlessly applied to 2D single-channel images without any modifications. However, in the case of multichannel images, we recommend reporting both of these metrics for each channel individually.
The $AP(i,p)$ metric can be effectively employed as long as a domain-agnostic approach to identify anatomical structures for a given organ or problem is available. This could involve training a segmentation model to identify anatomies across all the different sites for limited applications. 
\vspace{-0.2cm}
\section{Conclusion}
\vspace{-0.2cm}
\label{sec:conclu}
Given the increasing emphasis on medical image harmonization for deep learning applications, we propose a novel set of metrics $nWD(i,p)$, $nWD(t,p)$, and $AP(i,p)$.  These metrics are helpful in measuring the effectiveness of harmonization in the absence of rigorous gold standards for harmonization. Through various experiments, we demonstrate that these metrics perform equivalent to other already established image quality metrics and have unique interpretations. We anticipate that these image harmonization metrics will help benchmark the image harmonization performance in future studies.
\newpage
\noindent
\textbf{Compliance with ethical standards} \\
This research study was conducted retrospectively using human subject data. Ethical approval was received from the Institutional Review Board at Children’s National Hospital.

\vspace{4pt}
\noindent
\textbf{Acknowledgements} \\ 
This work was supported by The National Cancer Institute award UG3CA236536. 
\bibliographystyle{IEEEbib}
\bibliography{strings,refs}

\begin{thebibliography}{10}

\bibitem{mirza2014conditional}
M.~Mirza and S.~Osindero,
\newblock ``Conditional generative adversarial nets,''
\newblock {\em arXiv:1411.1784}, 2014.

\bibitem{sinha2021alzheimer}
Surabhi Sinha, Sophia~I Thomopoulos, Pradeep Lam, Alexandra Muir, and Paul~M Thompson,
\newblock ``Alzheimer’s disease classification accuracy is improved by mri harmonization based on attention-guided generative adversarial networks,''
\newblock in {\em 17th international symposium on medical information processing and analysis}. SPIE, 2021, vol. 12088, pp. 180--189.

\bibitem{medstyle}
Mengting Liu, Piyush Maiti, Sophia Thomopoulos, Alyssa Zhu, Yaqiong Chai, Hosung Kim, and Neda Jahanshad,
\newblock ``Style transfer using generative adversarial networks for multi-site mri harmonization,''
\newblock in {\em International Conference on Medical Image Computing and Computer-Assisted Intervention}. Springer, 2021, pp. 313--322.

\bibitem{liu_unit_2017}
M.-Y. Liu, T.~Breuel, and J.~Kautz,
\newblock ``Unsupervised image-to-image translation networks,''
\newblock in {\em NIPS'17}, 2017, p. 700–708.

\bibitem{carlos}
Carlos Tor-Diez, Antonio~Reyes Porras, Roger~J Packer, Robert~A Avery, and Marius~George Linguraru,
\newblock ``Unsupervised mri homogenization: application to pediatric anterior visual pathway segmentation,''
\newblock in {\em International Workshop on Machine Learning in Medical Imaging}. Springer, 2020, pp. 180--188.

\bibitem{yang_cyclegan_2018}
H.~Yang, J.~Sun, A.~Carass, C.~Zhao, J.~Lee, Z.~Xu, and J.~Prince,
\newblock ``Unpaired brain mr-to-ct synthesis using a structure-constrained cyclegan,''
\newblock in {\em Deep Learning in Medical Image Analysis and Multimodal Learning for Clinical Decision Support}, 2018, pp. 174--182.

\bibitem{harm_review}
Fengling Hu, Andrew~A. Chen, Hannah Horng, Vishnu Bashyam, Christos Davatzikos, Aaron Alexander-Bloch, Mingyao Li, Haochang Shou, Theodore~D. Satterthwaite, Meichen Yu, and Russell~T. Shinohara,
\newblock ``Image harmonization: A review of statistical and deep learning methods for removing batch effects and evaluation metrics for effective harmonization,''
\newblock {\em NeuroImage}, vol. 274, pp. 120125, 2023.

\bibitem{psnr}
Alain Horé and Djemel Ziou,
\newblock ``Image quality metrics: Psnr vs. ssim,''
\newblock in {\em 2010 20th International Conference on Pattern Recognition}, 2010, pp. 2366--2369.

\bibitem{denck2021mr}
Jonas Denck, Jens Guehring, Andreas Maier, and Eva Rothgang,
\newblock ``Mr-contrast-aware image-to-image translations with generative adversarial networks,''
\newblock {\em International Journal of Computer Assisted Radiology and Surgery}, vol. 16, pp. 2069--2078, 2021.

\bibitem{ssim}
Zhou Wang, Eero~P Simoncelli, and Alan~C Bovik,
\newblock ``Multiscale structural similarity for image quality assessment,''
\newblock in {\em The Thrity-Seventh Asilomar Conference on Signals, Systems \& Computers, 2003}. Ieee, 2003, vol.~2, pp. 1398--1402.

\bibitem{umap}
Leland McInnes, John Healy, and James Melville,
\newblock ``Umap: Uniform manifold approximation and projection for dimension reduction,''
\newblock {\em arXiv preprint arXiv:1802.03426}, 2018.

\bibitem{wasser}
{Peyre, R\'emi},
\newblock ``Comparison between w2 distance and 1 norm, and localization of wasserstein distance,''
\newblock {\em ESAIM: COCV}, vol. 24, no. 4, pp. 1489--1501, 2018.

\bibitem{kolouri2018sliced}
Soheil Kolouri, Phillip~E Pope, Charles~E Martin, and Gustavo~K Rohde,
\newblock ``Sliced-wasserstein autoencoder: An embarrassingly simple generative model,''
\newblock {\em arXiv preprint arXiv:1804.01947}, 2018.

\bibitem{fischl2012freesurfer}
Bruce Fischl,
\newblock ``Freesurfer,''
\newblock {\em Neuroimage}, vol. 62, no. 2, pp. 774--781, 2012.

\bibitem{zavaliangos2022testing}
Artemis Zavaliangos-Petropulu, Meral~A Tubi, Elizabeth Haddad, Alyssa Zhu, Meredith~N Braskie, Neda Jahanshad, Paul~M Thompson, and Sook-Lei Liew,
\newblock ``Testing a convolutional neural network-based hippocampal segmentation method in a stroke population,''
\newblock {\em Human Brain Mapping}, vol. 43, no. 1, pp. 234--243, 2022.

\bibitem{opfer2023automatic}
Roland Opfer, Julia Kr{\"u}ger, Lothar Spies, Ann-Christin Ostwaldt, Hagen~H Kitzler, Sven Schippling, and Ralph Buchert,
\newblock ``Automatic segmentation of the thalamus using a massively trained 3d convolutional neural network: higher sensitivity for the detection of reduced thalamus volume by improved inter-scanner stability,''
\newblock {\em European Radiology}, vol. 33, no. 3, pp. 1852--1861, 2023.

\bibitem{mnitemplate}
John Mazziotta, Arthur Toga, Alan Evans, Peter Fox, Jack Lancaster, Karl Zilles, Roger Woods, Tomas Paus, Gregory Simpson, Bruce Pike, et~al.,
\newblock ``A four-dimensional probabilistic atlas of the human brain,''
\newblock {\em Journal of the American Medical Informatics Association}, vol. 8, no. 5, pp. 401--430, 2001.

\end{thebibliography}

\end{document}